\documentclass[preprint,review,12pt]{elsarticle}
\usepackage{extsizes}

\usepackage[version=3]{mhchem}
\usepackage[left=1.5cm, right=1.5cm, top=1.785cm, bottom=2.0cm]{geometry}
\usepackage{times,mathptmx}
\usepackage{gensymb}
\usepackage{sectsty}
\usepackage{epsfig}
\usepackage{lastpage}
\usepackage{float}
\usepackage{fnpos}
\usepackage[english]{babel}
\usepackage{array}
\usepackage{droidsans}
\usepackage{charter}
\usepackage[T1]{fontenc}
\usepackage[usenames,dvipsnames]{xcolor}
\usepackage{setspace}
\usepackage[compact]{titlesec}
\usepackage{array}
\usepackage{color}
\usepackage{textcomp}
\usepackage{amsmath}
\usepackage{amssymb}
\usepackage{amsthm}

\usepackage{epstopdf}

\makeatletter
\def\ps@pprintTitle{%
     \let\@oddhead\@empty
     \let\@evenhead\@empty
     \def\@oddfoot{}%
		 \let\@evenfoot\oddfoot}
\makeatother

\begin{document}

\begin{frontmatter}

\title{Swelling of responsive-microgels: experiments versus models}

\author[DIP,ISC]{Valentina Nigro \footnote{Corresponding author: valentina.nigro@.uniroma1.it}}
\author[ISC,DIP]{Roberta Angelini}
\author[IPCF]{Monica Bertoldo}
\author[ISC,DIP]{Barbara Ruzicka}

\address[DIP]{Dipartimento di Fisica, Sapienza Universit$\grave{a}$ di Roma, P.le Aldo Moro 5, 00185 Roma, Italy.}
\address[ISC]{Istituto dei Sistemi Complessi del Consiglio Nazionale delle Ricerche (ISC-CNR), Sede Sapienza, Pz.le Aldo Moro 5, 00185 Roma, Italy}

\address[IPCF]{Istituto per i Processi Chimico-Fisici del Consiglio Nazionale delle Ricerche (IPCF-CNR), Area della Ricerca, Via G.Moruzzi 1, I-56124 Pisa, Italy.}

\begin{abstract}
Interpenetrated Polymer Network (IPN) microgels of PNIPAM and PAAc have been investigated and the experimental data have been compared with theoretical models from the Flory-Rehner theory. We confirm that the swelling behavior of PNIPAM microgels is well described by this theory by considering the second order approximation for the volume fraction $\phi$ dependence of the Flory parameter $\chi(\phi)$. Indeed the Volume-Phase Transition (VPT) of the PNIPAM-PAAc IPN microgel at neutral conditions and in $D_2O$ solvents can be well described only considering a third-order approximation. Interestingly we empirically find that sharper is the transition  higher is the order of the $\chi(\phi)$ relation which has to be considered.  Moreover the VPT can be experimentally controlled by tuning the polymer/solvent interactions through pH and solvent allowing to directly modify the delicate balance between energetic and entropic contributions and to explore the swelling behavior in a wide range of environmental conditions. In particular we find that the most advantageous condition for swelling is in water at acidic pH.
\\
\end{abstract}


\end{frontmatter}

\section{Introduction}
\label{Introduction}

Responsive microgels are aqueous dispersions of nanometre- or micrometre-sized hydrogel particles\cite{LyonRevPC2012, PaloliSM2012, PritiJCP2014}, able to swell and retain large amount of water in response to slight changes in the environmental conditions. The high responsiveness of this interesting class of smart materials makes responsive microgels attractive for several technological applications \cite{ VinogradovCurrPharmDes2006, DasAnnRevMR2006, ParkBiomat2013, HamidiDrugDeliv2008, SmeetsPolymSci2013, SuBiomacro2008} and very good model systems for understanding the complex behaviors emerging in soft colloids \cite{LikosJPCM2002,RamirezJPCM2009,HeyesJCP2009}. 
Indeed they allow to modulate the interparticle potential and their effective volume fraction through easily accessible control parameters such as temperature, pH or solvent, leading to novel phase-behaviors \cite{ WangChemPhys2014, PritiJCP2014, HellwegCPS2000, PaloliSM2012, WuPRL2003}, drastically different from those of conventional hard colloidal systems \cite{PuseyNat1986, ImhofPRL1995, PhamScience2002, EckertPRL2002, LuNat2008, RoyallNatMat2008, RuzickaNatMat2011, AngeliniNC2014}.

The most known responsive microgels are based on
 poly(N-isopropylacrylamide) (PNIPAM), which is a
 thermo-sensitive polymer \cite{PeltonColloids1986} with a Lower Critical Solution Temperature (LCST) in water at about 306 K. 
At room temperature indeed, the polymer is hydrophilic and strongly hydrated in solution, while it becomes hydrophobic above 306 K, where a coil-to-globule transition takes place. This gives rise to a Volume-Phase Transition (VPT) from a swollen to a shrunken state of any PNIPAM-based microgel~\cite{WuMacromol2003}. This typical swelling/shrinking behavior has been shown to be the driving mechanism of the phase behavior of aqueous suspensions of PNIPAM microgels, since it affects the interactions between particles and provides good tunability of both softness and volume fraction as a function of temperature \cite{MattssonNature2009, PaloliSM2012, LyonRevPC2012, PritiJCP2014, WangChemPhys2014}. 
It has been recently shown that the microgel
swelling/shrinking behavior can be strongly affected by
concentration \cite{TanPolymers2010,WangChemPhys2014},
 solvents \cite{ZhuMacroChemPhys1999} and
internal structure and composition (such as number and distribution of cross-linking
points \cite{KratzBerBunsenges11998, KratzPolymer2001} and core-shell structure~\cite{HellwegLangmuir2004, MengPhysChem2007}) or by introducing
additives into the PNIPAM network \cite{HellwegLangmuir2004}.

In this context PNIPAM microgels containing another specie as
co\textendash{}monomer or interpenetrated polymer are
even more interesting, as a more complex scenario comes out. In particular, addition of
poly(acrilic acid) (PAAc) to PNIPAM microgel provides an additional pH-sensitivity to the thermo-responsive microgel. 
Therefore their temperature dependent volume phase transition can be controlled by the content of AAc
monomer~ \cite{HuAdvMater2004, MaColloidInt2010}, by the pH \cite{KratzColloids2000, KratzBerBunsenges21998,
XiaLangmuir2004, JonesMacromol2000, NigroJNCS2015, NigroJCP2015} or the ionic strength \cite{KratzColloids2000, XiongColloidSurf2011} of the
suspension, since it strongly depends
on the effective charge density.

In this framework the synthesis procedure plays a crucial role, since the response of PNIPAM/PAAc microgels is strictly related 
to the mutual interference between the two monomers \cite{KratzColloids2000,
KratzBerBunsenges21998, JonesMacromol2000, XiongColloidSurf2011,
MengPhysChem2007, LyonJPCB2004, HolmqvistPRL2012, DebordJPCB2003}. In particular interpenetration of hydrophilic PAAc
into the PNIPAM microgel network (IPN PNIPAM-PAAc microgel) ~\cite{HuAdvMater2004, XiaLangmuir2004, XiaJCRel2005,
ZhouBio2008, XingCollPolym2010, LiuPolymers2012}, has little influence on the coil-to-globule transition temperature of the PNIPAM chains \cite{MaColloidInt2010}, leaving the temperature
dependence of the VPT almost unchanged with respect to the case of pure
PNIPAM microgel. Nevertheless the different solubility of PAAc at acidic and neutral pH, introduces an additional parameter to control the delicate balance between polymer/polymer and polymer/solvent interactions. At acidic pH the PAAc chains are not effectively solvated by water and H-bonds between the carboxylic (COOH-) groups of PAAc and the isopropyl (CONH-) groups of PNIPAM are favored~\cite{SibandMacromol2011, NigroJCP2015}; at neutral pH, instead, the balance between PNIPAM/PAAc and water/PAAc H-bonds is reversed: both compounds are therefore well solvated and water mediates their interaction, making the two networks completely independent one to each other. 

The typical swelling behavior of these thermo-responsive microgels has attracted great interest in the last years, with an increasing number of theoretical works aimed at describing this phenomenon \cite{Flory1953, LeleMacromol1998, HinoJAPS1996, OtakeMacromol1990, LopezLeonPRE2007}. In particular, for PNIPAM-based microgels it is well known that the swelling behavior can be well described by the classic Flory-Rehner theory \cite{Flory1953}, which has been extended to take into account the cross-linker influence on the polymer-solvent interaction \cite{CrassousCPS2008, HertleCPS2010, BalaceanuMacromol2013}. It has been shown that the driving force for swelling can be estimated from the properties of linear PNIPAM solutions, while the microgel elasticity opposing swelling is mainly due to the network topology dependent on the cross-linker concentration.
On the other hand, it has been shown that the swelling behavior is highly influenced by the effective charge density, which in PNIPAM-PAAc IPN microgels can be experimentally controlled by the content of AAc monomer~\cite{HuAdvMater2004, MaColloidInt2010} or by the pH of the suspension.

In this framework the swelling behavior of aqueous suspensions of PNIPAM-PAAc IPN microgels as a function of temperature, pH and concentration has been extensively investigated by our group through Dynamic Light Scattering (DLS) and Small-Angle Neutron Scattering (SANS) \cite{NigroJNCS2015, NigroJCP2015}. In particular it has been observed that, in the high dilution regime (where the interparticle interactions are negligible and phase separation does not occur), a VPT around 305-307 K occurs. However interesting differences depending on pH and H/D isotopic substitution have been found, suggesting that H-bonding plays a non-trivial role. 
Nevertheless there is a weak understanding of the real fundamental reasons of their peculiar behaviors. Therefore the aim of this work is to compare our experimental data with theoretical models accounting for the sharpness of the transition, in order to describe the swelling behavior of PNIPAM-PAAc IPN microgel, where the VPT can be experimentally controlled by tuning pH or acrylic acid content and by isotope substitution in the solvent.

\section{Experimental Methods}
\label{Experimental Methods}

\subsection{Sample preparation}
\paragraph{Materials}
N-isopropylacrylamide (NIPAM) from Sigma-Aldrich and
N,N'-methylene-bis-acrylamide (BIS) from Eastman Kodak were
recrystallized from  hexane and methanol, respectively. Acrylic acid (AAc)
from Sigma-Aldrich was purified by distillation (40 mmHg, 337 K). Sodium dodecyl sulphate (SDS),
98 \% purity, potassium persulfate (KPS), 98 \% purity, ammonium
persulfate, 98 \% purity, N,N,N\textasciiacute{},N\textasciiacute{}-tetramethylethylenediamine
(TEMED), 99 \% purity, ethylenediaminetetraacetic acid (EDTA),
NaHCO$_3$, were all purchased from Sigma-Aldrich and used as
received. Ultrapure water (resistivity: 18.2 M$\Omega$/cm at 298
K) was obtained with Sarium® pro Ultrapure Water purification
Systems, Sartorius Stedim from demineralized water. D$_2$O (99.9 atom \%) from Sigma-Aldrich.
All other solvents were RP grade (Carlo Erba) and were used as received.
A dialysis tubing cellulose membrane, HCWO 14000 Da,
from Sigma-Aldrich, was cleaned before use by washing in distilled water and NaHCO$_3$/EDTA solution.

\paragraph{Synthesis of PNIPAM and IPN microgels}
The IPN microgel was synthesized by a sequential  free radical
polymerization method~\cite{HuAdvMater2004}. In the first step PNIPAM micro-particles
were synthesized with (4.0850 $\pm$ 0.0001) g of NIPAM,
(0.0695 $\pm$ 0.0001) g of BIS and (0.5990 $\pm$ 0.0001) g of SDS in 300 mL of ultrapure water. The solution was purged with
nitrogen for 30 min, heated at (343.0 $\pm$ 0.3) K and the polymerization was initiated with
 (0.1780 $\pm$ 0.0001) g of KPS. After 4h the resulting
PNIPAM microgel was purified by dialysis against
distilled water for 2 weeks. For the synthesis of the second network, (65.45 $\pm$ 0.01) g of the
recovered PNIPAM dispersion (1.02 wt\% and 80 nm diamter at 298 K) was diluted with ultrapure water up to a volume of 320 mL.
(0.50 $\pm$ 0.01)g of BIS were mixed and the mixture was purged with nitrogen for 1 h. 2.3 mL of AAc
and (0.2016 $\pm$ 0.0001) g of TEMED and (0.2078 $\pm$ 0.0001) g of
ammonium persulfate were added and the 
reaction was carried out for 65 min at (295 $\pm$ 1)K. The obtained IPN microgel was purified by
dialysis against distilled water for 2
weeks, and then lyophilized to constant weight. The PAAc second network amount in the IPN was 31.5\% by gravimetric analysis and the acrylic acid monomer units content 22\% by acid/base titration. The remaining 9.5\% being ascribed to BIS.
The sample was lyophilized and redispersed again in H$_2$O or in D$_2$O by magnetic stirring for 1 day to obtain the final suspension at the required weight concentration.
Samples at different weight concentrations (wt.\%), in the following referred to simply as $C_w$, were obtained by dilution and by keeping constant the molar ratio in H$_2$O and D$_2$O. Therefore in the following we will refer to the weight concentration in H$_2$O.

 \subsection{DLS set-up and data analysis}
 \label{DLS set-up and data analysis}

DLS measurements have been performed with a multiangle light
scattering setup. The monochromatic and polarized beam  emitted
from a solid state laser (100 mW at $\lambda$=642 nm) is focused on the sample placed in a cylindrical VAT for index matching and temperature control. The scattered
intensity is simultaneously collected by single mode optical fibers at five different scattering
angles, namely $\theta=30\degree, 50\degree, 70\degree,
90\degree, 110\degree$, corresponding to different
scattering vectors $Q$, according to the relation
Q=(4$\pi$n/$\lambda$) sin($\theta$/2). 
In this way one can simultaneously measure 
the normalized intensity autocorrelation function
$g_2(Q,t)=<I(Q,t)I(Q,0)>/<I(Q,0)>^{2}$ at five different $Q$ values
with a high coherence factor close to the ideal unit value.
Measurements have been performed as a function of temperature
in the range T=(293$\div$313) K across the VPT at low weight concentrations, at both acidic and
neutral pH. Reproducibility has been tested by repeating
measurements several times. 

\begin{figure}
\begin{centering}
\includegraphics[width=8cm]{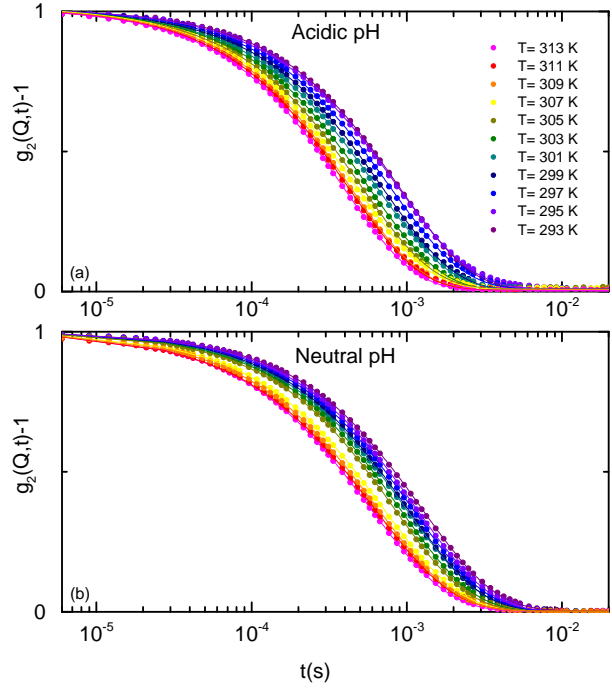}
\par\end{centering}
\caption{\label{fig:corr func}Normalized intensity autocorrelation
functions of an IPN sample at $C_w=0.10$ \%, at (a) acidic and (b) neutral pH and
$\theta$=90\textdegree ~for the indicated temperatures. Full lines are fits according to Eq.(\ref{Eqfit})}
\end{figure}

In Fig.~\ref{fig:corr func} the typical behavior of the normalized
intensity autocorrelation functions for an IPN sample at weight
concentration $C_w$=0.10 \%, at acidic ($\approx 5$) and neutral ($\approx 7$) pH and $\theta$=90\degree~ for
the indicated temperatures is shown.

As commonly known, the intensity correlation function of most
colloidal systems is well described by the Kohlrausch-Williams-Watts
expression~\cite{KohlrauschAnnPhys1854, WilliamsFaradayTrans1970}:

\begin{equation}
g_2(Q,t)=1+b[(e^{-t/\tau})^{\beta}]^{2} \label{Eqfit}
\end{equation}

where $b$ is the coherence factor, $\tau$ is an "effective"
relaxation time and $\beta$ describes the deviation from the simple exponential decay ($\beta$
= 1) usually found in monodisperse systems and gives a measure of the distribution of relaxation times. Many glassy materials show a stretching of the correlation functions (here referred to as "stretched behavior") characterized by an exponent $\beta$< 1.  
In the case of Brownian diffusion it is possible to relate the relaxation time to the translational
diffusion coefficient $D_t$ through the relation
$\tau=1/(Q^2D_t)$. In the limit of non interacting spherical
particles the hydrodynamic radius is given by the Stokes Einstein relation

\begin{equation}
R=K_B T /6 \pi \eta D_t \label{StokesEinstein}
\end{equation}

where $\eta$ is the sample viscosity, $T$ is the sample temperature and $k_B$ is the Boltzmann constant.
Indeed, once the viscosity $\eta$ is known, this relation allows to calculate the hydrodynamic
radius from the diffusion coefficient.
In particular the hydrodynamic radii of this work have been calculated by using the viscosity of the solvent, since samples are in the high dilution limit.

\section{Swelling Theory}

Responsive microgels based on thermo-sensitive polymers, such as those investigated in this work, exhibit a complex and exotic behavior driven by the response to temperature changes. 
It is well known that the VPT of microgel particles is closely related to the coil-to-globule transition of the polymer chains composing the microgel network and it has been extensively explained by the Flory-Rehner theory \cite{Flory1953}. At equilibrium the net osmotic pressure in a microgel, consisting of a mixing $\pi_{mixing}$ and an elastic $\pi_{elasticity}$ contribution, is equal to zero

\begin{equation}
\pi = \pi_{mixing} + \pi_{elasticity}= 0
\end{equation}

This implies that \cite{Flory1953}

\begin{equation}
ln(1-\phi)+\phi+\chi \phi^2+\frac{\phi_0}{N}[(\frac{\phi}{\phi_0})^{1/3}-\frac{1}{2}\frac{\phi}{\phi_0}]=0
\label{EqState}
\end{equation}

\noindent where $\chi$ is the Flory polymer-solvent energy parameter (empirically given as a function of temperature and microgel weight fraction), $N$ is the average number of segments between two neighboring cross-linking points in the microgel network, $\phi$ is the polymer volume fraction within the particle and $\phi_0$ is the polymer volume fraction in the reference state, where the conformation of the network chains is the closest to that of unperturbed Gaussian chains. Typically this reference state is taken as the shrunken one and $\phi_0$ is the polymer volume fraction corresponding to the shrunken state. 
For isotropic swelling, $\phi$ can be related to the particle size $R$ as
\begin{equation}
\frac{\phi}{\phi_0}=(\frac{R_0}{R})^{3}
\label{phi}
\end{equation}

\noindent where $R_0$ is the particle diameter at the reference state (i.e. the shrunken state) and $R$ is the particle diameter at a given state. 

Therefore from Flory-Rehner theory we obtain the relation between the particle size and the other relevant parameter, thus giving details about the thermodynamics of the swelling/deswelling mechanism.
In this framework the main role is played by the Flory $\chi$ parameter, reflecting the influence of the energy and entropy changes during the mixing process. It is indeed defined as the free energy change per solvent molecule when both polymer-polymer and solvent-solvent contacts are replaced by polymer-solvent contacts  \cite{Flory1953, LopezLeonPRE2007, ShibayamaMacromol1997}

\begin{equation}
\chi=\frac{\Delta G}{k_B T}=\frac{\Delta H - T\Delta S}{k_B T}=\frac{1}{2}-A(1-\frac{\theta}{T})
\label{Chi}
\end{equation}

\noindent where $k_B$ is the Boltzmann constant, $T$ is the temperature and $A$ and $\theta$ are defined as

\begin{equation}
A=\frac{2\Delta S+k_B}{2 k_B}
\label{A}
\end{equation}

\begin{equation}
\theta=\frac{2\Delta H}{2\Delta S+k_B}
\label{Theta}
\end{equation}

A rich behavior is observed depending on $\chi$, which embodies the temperature dependence of the swelling behavior. For $T=\theta$, the Flory parameter is $\chi=1/2$, corresponding to the equilibrium conditions: polymer coils act like ideal chains and the excess of osmotic pressure of mixing between polymer and solvent is zero. For values $\chi<1/2$, the solvent is a good solvent for the polymer, interactions between polymer segments and solvent molecules are energetically favorable and allow polymer coils to expand. In these conditions the mixing contribution to the osmotic pressure is positive, thus promoting swelling. For $\chi>1/2$ the solvent is a poor solvent, hence polymer-polymer self-interactions are preferred and the polymer coils contract, thus promoting deswelling.

However the original Flory-Rehner theory is not able to describe the discontinuous transition observed in soft colloidal microgels due to the high complexity of their structure. It has been shown \cite{ErmanMacromol1986,LopezLeonPRE2007} that exchange interactions must be described by accounting for interactions of higher order than contacts between molecules. This means that the $\chi$ parameter as a function of the volume fraction $\phi$, has to be written as a power series expansion
\begin{equation}
\chi=\chi_1 (T)+\chi_2 \phi+\chi_3 \phi^2+\chi_4 \phi^3+\cdots
\label{Chi3}
\end{equation}
where $\chi_1$ is the zero-order original Flory parameter of Eq.(\ref{Chi}) and $\chi_2,\chi_3,\chi_4,\ldots$ are temperature independent coefficients which introduce additional terms in the equation of state (Eq.(\ref{EqState})). 

This empirical model has been extensively used for describing the swelling behavior of homopolymeric PNIPAM microgels \cite{ShibayamaMacromol1997, LopezLeonPRE2007}. However we propose to apply the same theory to non-homogeneous systems, such as PNIPAM-PAAc IPN microgel, by assuming that $\chi$ is an effective mean parameter accounting for polymer/polymer interactions within each network, polymer/polymer interactions between different networks and polymer/solvent interactions.

\section{Results and Discussions}
\label{Results}

\begin{figure}
\centering
\includegraphics[height=8cm]{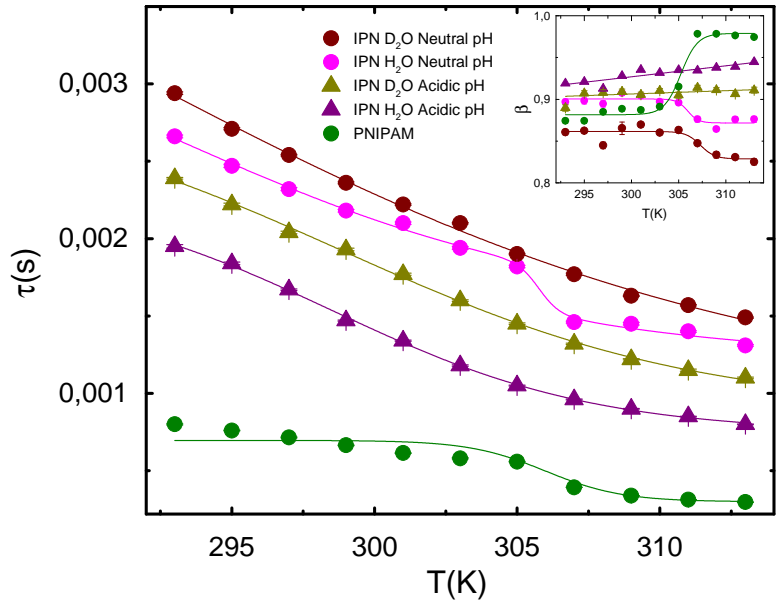}
\caption{Relaxation time and stretching parameter (inset) as a function of temperature for H$_2$O and D$_2$O suspensions of IPN microgels at acidic and neutral pH, compared with aqueous suspensions of PNIPAM microgels at the same weight concentration ($C_w=0.10 \%$). Full lines are guides for eyes.}
\label{fig:tauT}
\end{figure}

In Fig.\ref{fig:tauT} we report the comparison between the temperature dependence of the relaxation time, as obtained through a fit with Eq.(\ref{Eqfit}), for D$_2$O and H$_2$O suspensions of IPN microgels, at both acidic and neutral pH and for aqueous suspension of PNIPAM microgels, at the same fixed concentration ($C_w=0.10 \%$).
In all the investigated samples a dynamical transition associated to the VPT, from a swollen to a shrunken state, is evidenced \cite{NigroJNCS2015}. The relaxation time slightly decreases as temperature increases, until the transition is approached around T=305 K. Above this Volume-Phase Transition Temperature (VPTT) it decreases to its lowest values, corresponding to the shrunken state. 

However interesting differences between pure PNIPAM microgels and PNIPAM-PAAc IPN microgels can be noticed.
In particular, the VPT of IPN microgels is strongly affected by the pH of the solution, due to the different solubility of PAAc at acidic and neutral pH. A first evidence of this pH-sensitivity is observed in the temperature behavior of the normalized intensity autocorrelation functions (Fig.\ref{fig:corr func}), since an almost continuous transition at acidic pH (Fig.\ref{fig:corr func}(a)) and a sharp transition at neutral pH (Fig.\ref{fig:corr func}(b)) are observed around 305-307 K. Indeed as temperature increases, the relaxation time slowly decreases up to the VPTT, while above this temperature different behaviors come out depending on pH and solvent. For H$_2$O samples at acidic pH an almost continuous transition is observed. At neutral pH, instead, the H-bonds interactions between the carboxylic groups of PAAc and PNIPAM are reduced and a sharp transition to lower values of $\tau$ is observed. 
Moreover, although the main features of the dynamical behavior for H$_2$O suspensions of PNIPAM-PAAc IPN microgels \cite{NigroJNCS2015} are preserved under isotopic substitution, some differences can be observed in D$_2$O samples: the sharpness of the transition is widely reduced in D$_2$O with respect to H$_2$O suspensions at neutral pH and the relaxation times are always higher in D$_2$O than in H$_2$O, suggesting a slowing down of the dynamics under isotopic substitution, probably due to the higher viscosity of D$_2$O compared to H$_2$O.
Moreover the stretching parameter $\beta$ (inset of Fig.\ref{fig:tauT}) provides useful details on samples polydispersity. Indeed for PNIPAM microgels $\beta \approx 0.9$ indicates slightly stretched correlation curves and an almost monodisperse sample. For IPN microgels the scenario is more complex: at acidic pH $\beta \approx0.9$ and the sample is mainly monodisperse in all the investigated temperature range, while at neutral pH it decreases with temperature up to $\beta \approx 0.8$, indicating more polydisperse samples above the VPTT.

\begin{figure}
\begin{centering}
\includegraphics[width=8cm]{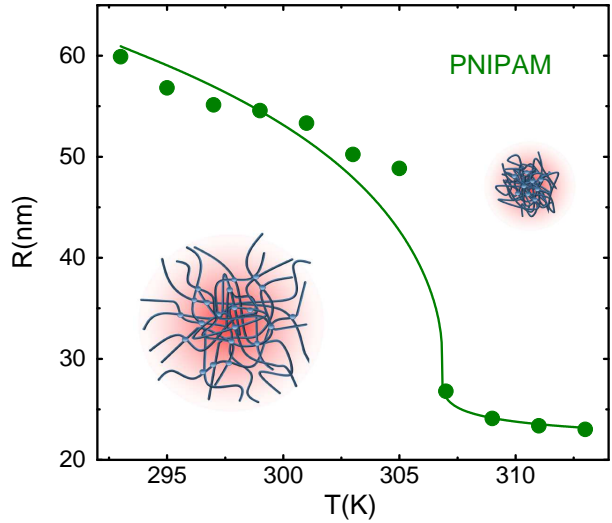}
\par\end{centering}
\caption{\label{fig:radiusPNIPAM} Temperature behavior of the hydrodynamic radius obtained through DLS measurements for aqueous suspensions of PNIPAM microgels  at $C_w$=0.10 \%. Solid line is the fit obtained through Eq.(\ref{EqTR}).}
\end{figure}
 
All these features can be well visualized by looking at the temperature behavior of the hydrodynamic radius, as obtained from DLS measurements in the high dilution limit, where interparticle interactions can be neglected and the Stokes-Einstein relation (Eq.(\ref{StokesEinstein})) can be used.
In Fig.\ref{fig:radiusPNIPAM} the temperature behavior of the hydrodynamic radius for aqueous suspensions of PNIPAM microgels at weight concentration C$_w$=0.10 \% is shown. A cross-over from a swollen to a shrunken state is observed at T$\approx$306 K. 
It is well known that the swelling behavior of PNIPAM microgels can be described by the Flory-Rehner theory, which well characterizes a discontinuous transition by assuming the volume fraction dependence of the Flory $\chi$ parameter (Eq.(\ref{Chi})). According to previous studies for PNIPAM microgels \cite{LopezLeonPRE2007}, a good agreement between theory and experiments is obtained with a second-order approximation of the $\chi(\phi)$ relation (Eq.(\ref{Chi})). 
Therefore by inverting Eq.(\ref{EqState}) it is possible to explicit the temperature dependence on the particle radius

\begin{equation}
T=\frac{-A \phi_0^2(\frac{R_0}{R})^6 \theta}{ln[1-\phi_0(\frac{R_0}{R})^3]+\phi_0(\frac{R_0}{R})^3+\phi_0^2(\frac{R_0}{R})^6(\frac{1}{2}-A)+\chi_2 \phi_0^3(\frac{R_0}{R})^9+\chi_3 \phi_0^4(\frac{R_0}{R})^{12}+\frac{\phi_0}{N}[\frac{R_0}{R}-\frac{1}{2}(\frac{R_0}{R})^3]}
\label{EqTR}
\end{equation}

The solid line in Fig.\ref{fig:radiusPNIPAM} is the best fit of our experimental data through Eq.(\ref{EqTR}),  with $\phi_0 \approx 0.8$, returning realistic fitting parameters. 
In particular the value of the Flory temperature $\theta=(306.7 \pm 0.3)~K$ is close to the VPTT, thus confirming that the volume phase transition is strictly related to the transition from good to poor solvent. Furthermore we find negative values of $A$, leading to negative values of $\Delta S$ and $\Delta H$, as expected for systems with a LCST and confirming that for PNIPAM microgels the transition in the solvent quality is driven by order-disorder processes due to hydrophobic interactions \cite{LopezLeonPRE2007}.

\begin{figure}
\begin{centering}
\includegraphics[width=8cm]{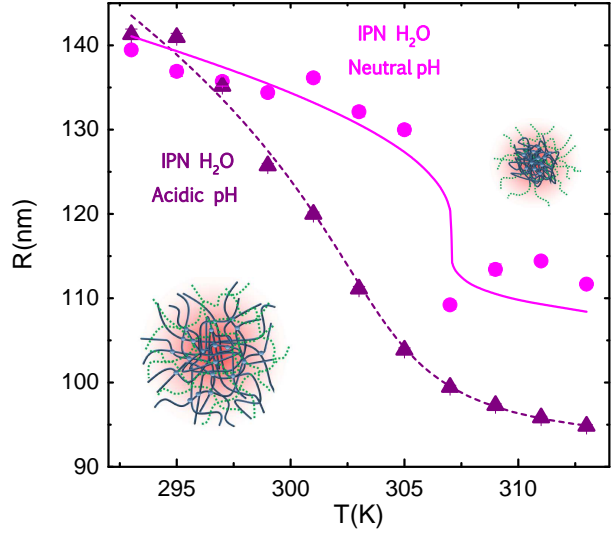}
\par\end{centering}
\caption{\label{fig:radiusIPNH2O} Temperature behavior of the hydrodynamic radius obtained through DLS measurements for H$_2$O suspensions of IPN microgels  at $C_w$=0.10 \%. Dashed line is the fit obtained through Eq.(\ref{EqTR}) with the second-order approximation of the $\chi$ parameter and solid line is the fit obtained through Eq.(\ref{EqTR2}) with the third-order approximation of the $\chi$ parameter.}
\end{figure}

As the acrylic acid is introduced within the PNIPAM network, the swelling capability is reduced with respect to PNIPAM microgels and an additional control parameter is introduced. In Fig.\ref{fig:radiusIPNH2O} the temperature behavior of the hydrodynamic radius for aqueous suspensions of IPN microgels at acidic and neutral pH is reported. The presence of PAAc in IPN microgels allows to modulate the sharpness and the amplitude of the VPT by changing pH or, equivalently, by deprotonating the sample. In particular at neutral pH the transition is sharper and narrower with respect to acidic pH, where a smoother and wider transition is observed, indicating a more hydrophobic system \cite{NigroJNCS2015}.  
This behavior can be interpreted in terms of the role played by the PAAc in IPN microgels \cite{NigroJCP2015}. At low pH the acrylic acid is insoluble in water and  is mainly in protonated state, therefore H-bonds between its carboxylic (COOH-) groups and the amidic (CONH-) groups of PNIPAM are formed and IPN microgels are strongly hydrophobic. Therefore when heated above the VPTT, due to the increased hydrophobic interactions, IPN microgels expel a large amount of water giving rise to a very dense shrunken state. At neutral pH the carboxylic groups of PAAc are deprotonated, leading to a strong charge repulsion which limits the formation of H-bonds with PNIPAM so that the two networks result independent and the sharpness of the transition is partially restored with respect to the case of pure PNIPAM.
 
Solid and dashed lines in Fig.\ref{fig:radiusIPNH2O} are the best fits of our experimental data, with $\phi_0 \approx 0.8$.
In particular Eq.(\ref{EqTR}) well describes our data only at acidic pH, where the transition is smooth and the $\phi$ dependence of the Flory $\chi$ parameter is well approximated by the second-order approximation of Eq.(\ref{Chi3}). In this case the value of the Flory temperature $\theta=(304.8 \pm 0.2)~K$ is close to the VPTT for IPN microgels at acidic pH \cite{NigroJNCS2015}, which is expected to be slightly shifted backward with respect to pure PNIPAM microgels. Furthermore we find negative values of $A$, indicating that the fundamental mechanism of the mixing process is not affected as the acrylic acid is introduced. These results confirm that at the IPN microgel structure can be modeled as an effective average network, indicating that the PNIPAM and PAAc networks are tightly bound to each other.

At neutral pH instead the VPT becomes sharp and discontinuous and the description of the data through Eq.(\ref{EqTR}) fails, suggesting that a second-order approximation of the $\chi$ parameter is no more sufficient. We decided therefore to extend Eq.(\ref{EqTR}) considering the third-order expansion of the $\chi(\phi)$ relation (Eq.(\ref{Chi3})):  

\begin{equation}
\begin{split}
T&=\\
&\frac{-A \phi_0^2(\frac{R_0}{R})^6 \theta}{ln[1-\phi_0(\frac{R_0}{R})^3]+\phi_0(\frac{R_0}{R})^3+\phi_0^2(\frac{R_0}{R})^6(\frac{1}{2}-A)+\chi_2 \phi_0^3(\frac{R_0}{R})^9+\chi_3 \phi_0^4(\frac{R_0}{R})^{12}+\chi_4 \phi_0^5(\frac{R_0}{R})^{15}+\frac{\phi_0}{N}[\frac{R_0}{R}-\frac{1}{2}(\frac{R_0}{R})^3]}
\label{EqTR2}
\end{split}
\end{equation}

By using this relation for fitting our experimental data, we find a good agreement between theory and experiments as shown in Fig.\ref{fig:radiusIPNH2O} at neutral pH. In particular we obtain values of the Flory temperature $\theta=(306.0 \pm 0.2)~K$ close to the VPTT for neutral pH samples, which is expected to be shifted to higher temperature with respect to acidic pH \cite{NigroJNCS2015, NigroJCP2015}. Moreover the obtained values of $A$ are negative and lower than those found at acidic pH, confirming that mixing is governed by the LCST and suggesting that the entropic change during the mixing process is less significant than at acidic pH.

\begin{figure}
\begin{centering}
\includegraphics[width=8cm]{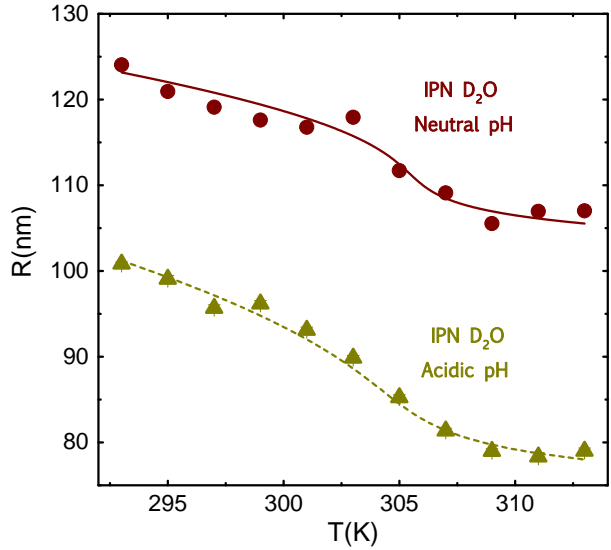}
\par\end{centering}
\caption{\label{fig:radiusIPND2O} Temperature behavior of the hydrodynamic radius obtained through DLS measurements, for D$_2$O suspensions of IPN microgels  at $C_w$=0.10 \%. Solid and dashed lines are the fit obtained through Eq.(\ref{EqTR2}) with the third-order approximation of the $\chi$ parameter.}
\end{figure}

In addition the swelling behavior is slightly affected by H/D isotopic substitution in the solvent, since H-bondings play a crucial role. The temperature dependence of the hydrodynamic radii of D$_2$O suspensions of IPN microgels at both acidic and neutral pH and at the same weight concentration ($C_w=0.10$ \%), is shown in Fig.\ref{fig:radiusIPND2O}. The presence of the acrylic acid  reduces the swelling capability of the microgel particles also in deuterated suspensions~\cite{HuAdvMater2004, XiaLangmuir2004, JonesMacromol2000}. Nevertheless for D$_2$O suspensions at acidic pH the transition appears slightly sharper and the range of variability of $R$ reduced with respect to H$_2$O suspensions, while at neutral pH the transition is smoother and the range of variability of $R$ unchanged. This leads to no significant changes at acidic or neutral pH, in contrast with H$_2$O suspensions where the swelling behavior is highly sensitive to pH. This suggests that the balance between polymer/polymer and polymer/solvent interactions strictly depends on the solvent and that in D$_2$O less evident changes with pH are expected in the solvent quality.
In this case a good agreement between theory and experiments is obtained with a third-order approximation of the $\chi(\phi)$ relation at both acidic and neutral pH. Solid and dashed lines in Fig.\ref{fig:radiusIPND2O} are the best fits of our experimental data through Eq.(\ref{EqTR2}),  with $\phi_0 \approx 0.8$.  
We obtain values of the Flory temperature $\theta=(304.9 \pm 0.4)~K$ at acidic pH and $\theta=(305.0 \pm 0.3)~K$ at neutral pH, as expected from previous studies on these samples \cite{NigroJNCS2015, NigroJCP2015}. Moreover the obtained values of $A$ are close to each other, thus confirming that in D$_2$O samples pH does not highly affects the swelling mechanism.

\begin{figure}
\begin{centering}
\includegraphics[width=8cm]{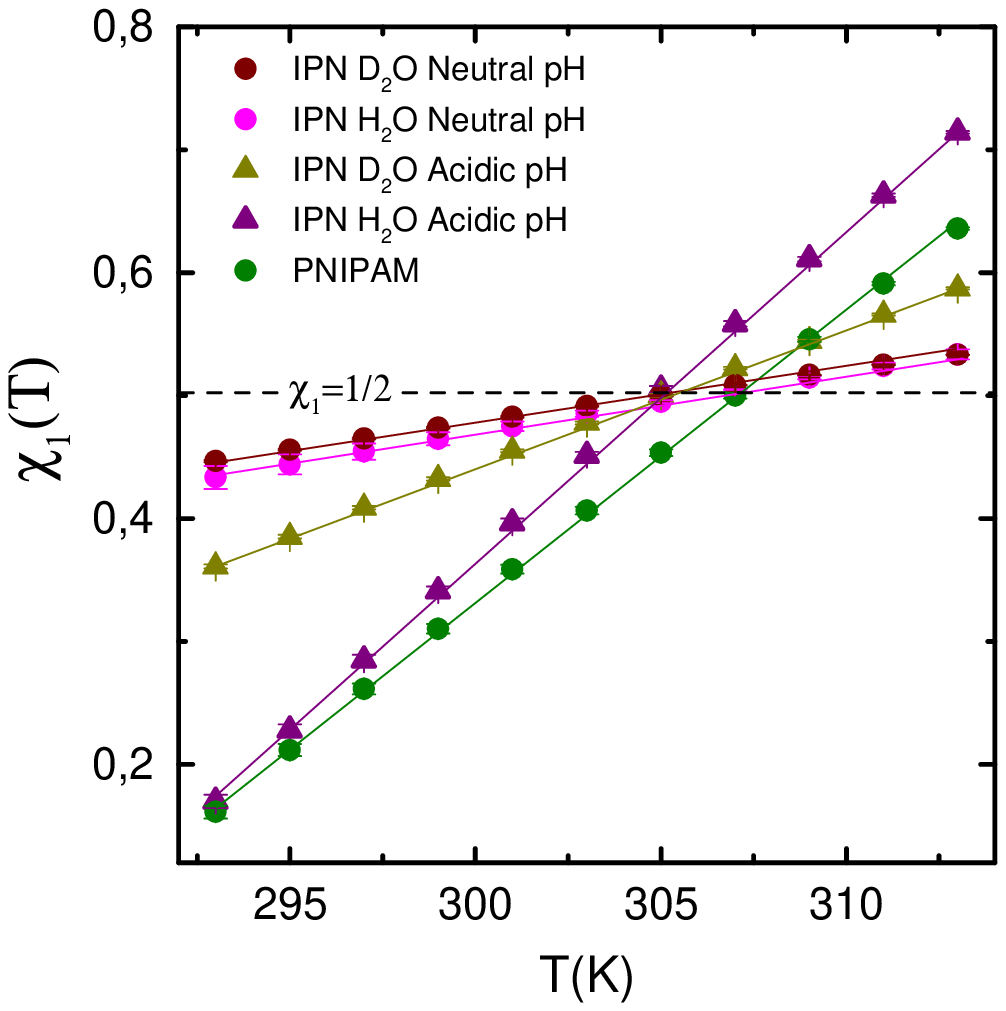}
\par\end{centering}
\caption{\label{fig:Chi1} Temperature behavior of the Flory $\chi_1$ parameter for PNIPAM microgels and IPN microgels in H$_2$O and D$_2$O, at both acidic and neutral pH. Solid lines are guides to eyes.}
\end{figure} 

\begin{table}
\fontsize{10}{7}
\center
\renewcommand\arraystretch{1}
\begin{tabular}{| l c  c  c |} \hline

{\newline \em \textbf{sample} \newline} & {\newline \newline} & {\newline \em \textbf{A} \newline} & {\newline \em \textbf{$\theta$}(K)\newline }\\ \hline \hline

\footnotesize PNIPAM &   & $-7.09 \pm 0.06$ & $306.7 \pm 0.3$ \\
\footnotesize IPN H$_2$O pH 5 &   & $-8.19 \pm 0.06$ & $304.8 \pm 0.2 $\\
\footnotesize IPN H$_2$O pH 7 &   & $-1.5 \pm 0.2$ & $ 306.2 \pm 0.2 $ \\ 
\footnotesize IPN D$_2$O pH 5 &   &$-3.40 \pm 0.03$ & $304.9 \pm  0.4$\\
\footnotesize IPN D$_2$O pH 7 &   &$-1.27 \pm 0.07$ & $ 305.0 \pm 0.5 $\\ \hline
\end{tabular}
\caption{\label{AT} Values of the \textit{A} parameter and Flory temperature $\theta$ as obtained from the best fit of the temperature behaviors of the hydrodynamic radii with Eq.(\ref{EqTR}) or Eq.(\ref{EqTR2}) depending on the sample.}
\end{table}

The differences in the swelling behaviors for the investigated samples can be well summarized by looking at the temperature behavior of the Flory $\chi_1$ parameter. Indeed the $\chi_1$ parameter reflects the influence of the energy and the entropy changes during the mixing process. It is the zero-order approximation of Eq.(\ref{Chi3}) and is therefore given by Eq.(\ref{Chi}), returning details about the delicate balance between energetic and entropic contributions. In Fig.\ref{fig:Chi1} the behavior of $\chi_1$ is reported for all the investigated samples. The obtained values of $A$ and $\theta$, reported in Tab.\ref{AT}, are such that $0<\chi_1<1$ as expected. Moreover we find $\chi_1<1/2$ below the VPTT and $\chi_1>1/2$ above the VPTT, thus indicating a transition from good to poor solvent across the volume-phase transition.
However a different slope characterizes the $\chi_1$ temperature behavior for different samples depending on pH and solvent, corresponding of the fitting parameters $A$ and $\theta$. In particular we find the largest slope for IPN microgels in water at acidic pH, indicating that mixing between polymer and solvent is favored and thus swelling promoted. This slope is even higher than for PNIPAM microgels, suggesting that the higher hydrophobicity due to the presence of PAAc may promote swelling. On the other hand the smallest slope of IPN microgels in water is found at neutral pH, meaning that the entropy changes are the least significant. Therefore the swelling process in H$_2$O is more favored at acidic than at neutral pH, being the system more hydrophobic. Moreover the $\chi_1$ behavior also depend on solvent: in D$_2$O the changes between the $\chi_1$ behaviors at acidic and neutral pH are less significant than in H$_2$O, confirming that the balance between polymer/polymer and polymer/solvent interactions strictly depends on the solvent and therefore on the H-bondings. 
\\
\begin{figure}
\begin{centering}
\includegraphics[width=14cm]{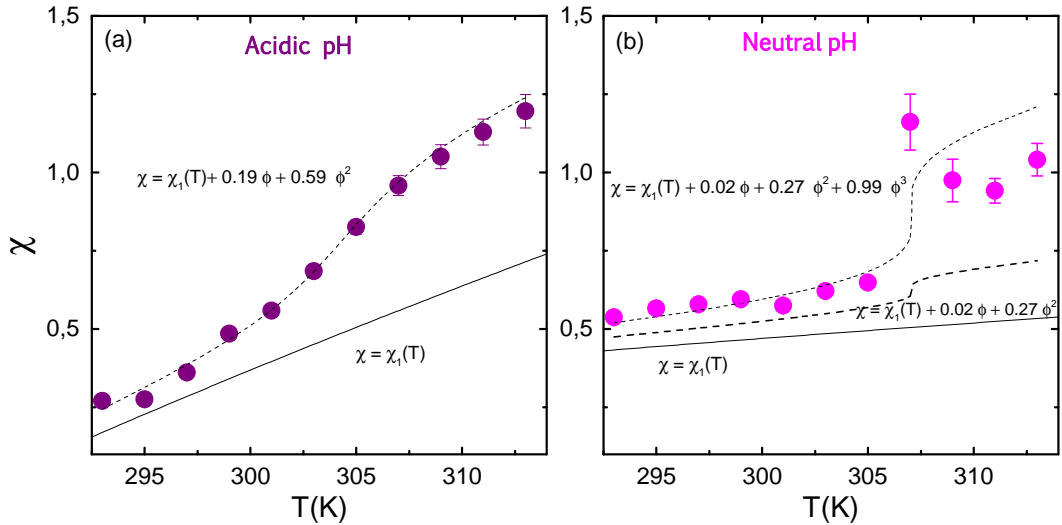}
\par\end{centering}
\caption{\label{fig:Chi} Temperature behavior of the $\chi$ parameter for H$_2$O suspensions of IPN microgels at (a) acidic and (b) neutral pH. Solid lines correspond to the Flory $\chi_1$ parameter, while dashed and dotted lines correspond to the theoretical predictions for $\chi(\phi)$.}
\end{figure}

From the temperature behavior of the hydrodynamic radius we experimentally calculate the $\chi$ parameter, which is plotted in Fig.\ref{fig:Chi} for H$_2$O samples as a function of T at acidic and neutral pH. At acidic pH (Fig.\ref{fig:Chi}(a)) an almost continuous transition from the swollen to the shrunken state is observed and a second-order approximation of $\chi(\phi)$ well describes the microgel swelling. On the other hand at neutral pH the second-order approximation of the $\chi(\phi)$ relation (dashed lines in Fig.\ref{fig:Chi}(b)) does not well reproduce the amplitude of the transition due to the increasing complexity of the microgel structure. Therefore a stronger $\phi$ dependence of $\chi$ is necessary to reproduce the sharp change across the VPT and a third-order approximation has to be taken into account (dotted line in Fig.\ref{fig:Chi}(b)) to empirically capture the essential physics of the many-body interactions, whose exact nature is still lacking. However at low temperature the experimental points are well described with the theoretical solid line corresponding to the $\chi_1(T)$ behavior, as previously observed for PNIPAM microgels \cite{LopezLeonPRE2007, ShibayamaMacromol1997}, even if the discrepancy between theoretical prediction and experimental points may suggest that particles are not fully swollen in the investigated temperature range.  
Therefore the Flory-Rehner theory well describes the main thermodynamic aspects of the volume-phase transition of PNIPAM-PAAc IPN microgels. However we empirically observe that a higher-order approximation of the $\chi(\phi)$ relation has to be considered as sharper is the transition. This is evident for IPN microgels in H$_2$O at neutral pH and for IPN microgels in D$_2$O at both acidic and neutral pH, where the discontinuous volume phase transition may be explained in terms of an increasing complexity of the microgel structure due to the breaking of the H-bonds betwen PNIPAM and PAAc networks. Therefore a third-order approximation of $\chi(\phi)$ is required to account for the balance between polymer/polymer and polymer/solvent interactions.

\section{Conclusions}
\label{Conclusions}

The swelling behavior of multi-responsive IPN microgels has been investigated and experimental data, obtained through DLS measurements, have been compared with  models from the swelling theory. The discontinuous transitions can be described by introducing the dependence of the original Flory $\chi$ parameter from the volume fraction $\phi$. In particular the swelling behavior of pure PNIPAM microgels is well described with a second-order approximation of the $\chi(\phi)$ relation, confirming previous results on these systems \cite{LopezLeonPRE2007}. Also for PNIPAM-PAAc IPN microgels at acidic pH an almost continuous transition from a swollen to a shrunken state is observed and a good agreement between theory and experiments is obtained with a second-order approximation of the $\chi(\phi)$, reflecting the almost homogeneous structure of the microgel network due to the strong interactions between the PNIPAM and the PAAc networks. On the contrary at neutral pH the transition becomes sharp and discontinuous and a higher, third-order approximation of the $\chi(\phi)$ relation has to be considered to account for the non-homogeneity of the system when polymer/polymer interactions are replaced by polymer/solvent interactions.  The same approximation is necessary to describe the swelling behavior of D$_2$O suspensions at both acidic and neutral pH. Indeed the VPT is affected by H/D isotopic substitution in the solvent, since the balance between polymer/polymer and polymer/solvent strictly depends on H-bondings. This leads to no significant changes with pH in D$_2$O suspensions and to an intermediate sharpness of the VPT with respect to H$_2$O samples at both acidic and neutral pH. Indeed H-bond interactions are weaker in D$_2$O than in H$_2$O and at both acidic and neutral pH a higher degree of complexity has to be introduced.  
Details about the delicate balance between energetic and entropic contributions are embodied in the Flory $\chi_1$ parameter. Its temperature behavior confirms that both for water and for heavy water the swelling is promoted at acidic pH, while at neutral pH the least advantageous condition for swelling is found. Therefore the observed behavior of $\chi_1(T)$ suggests that the swelling can be promoted or inhibited by controlling the polymer/solvent interactions through pH and solvent.

%
\subsection*{Bibliography}
%

\bibliographystyle{unsrt}\biboptions{sort&compress}

\end{document}